\documentclass[prl,aps,twocolumn,showkeys,preprintnumbers,showpacs,superscriptaddress]{revtex4}

\usepackage{epsfig,amssymb,amsfonts,verbatim}

\def\bfl{\begin{flushleft}}
\def\efl{\end{flushleft}}
\def\bfr{\begin{flushright}}
\def\efr{\end{flushright}}
\def\bc{\begin{center}}
\def\ec{\end{center}}
\def\be{\begin{equation}}
\def\ee{\end{equation}}
\def\ba{\begin{eqnarray}}
\def\ea{\end{eqnarray}}
\def\baa#1{\begin{array}{#1}}
\def\eaa{\end{array}}
\def\bw{\begin{widetext}}
\def\ew{\end{widetext}}
\def\nn{\nonumber }
\def\lb#1{\label{#1}}

\def\text#1{\mbox{#1}}

\begin{document}


\title{Phenomenological models of dielectric functions and screened Coulomb potential}

\author{Andrew Das Arulsamy}

\address{Condensed Matter Group, Division of Exotic Matter, No. 22, Jalan Melur 14,
Taman Melur, 68000 Ampang, Selangor DE, Malaysia}


\date{\today}



\begin{abstract}

The evolution of static dielectric constants and polarization with
doping are analyzed and discussed using established experimental
results. The variation of screened Coulomb potential with doping
is derived theoretically to justify the said evolutions. The
latter justification arises due to the influence of ionization
energy ($\mathcal{E_I}$) on the static dielectric functions. The
bare Coulomb potential and the Thomas-Fermi screened potential
with finite carrier density were recovered when $\mathcal{E_I}
\rightarrow \infty$ and $\mathcal{E_I} \rightarrow E_F^0$
respectively. Basically, these phenomenological models are
associated with the ionization energy based Fermi-Dirac statistics
(iFDS).

\end{abstract}


\maketitle



\section{1. Introduction}\lb{s-in}

Analogous to ferro- and paramagnetic systems that deal with spin
alignments, \verb"Ferroelectrics" (FE) exhibit spontaneous
polarization below the Curie temperature ($T_C$) whereas the
unpolarized state exists above $T_C$. The simplest
\verb"Ferroelectric" (FE) crystal, SnTe with NaCl structure was
reported by Riedl {\it et al}.~\cite{riedl1} back in 1965. The
\verb"Ferroelectricity" (FE) observed in SnTe is of displacive
nature, identical with other Perovskite FE. Presently, the
Perovskite structured (Ba,Sr)TiO$_4$ (BST) plays an enormous role
for the development of technological applications due to its high
spontaneous polarization, $\sim$10$^{-2}$ Cm$^{-2}$ at $T_C$
$\approx$ 290 K~\cite{kimura}. Although substituting Nb into the
BST system reduces the twin formation and $T_C$ significantly, its
dielectric constant and spontaneous polarization was found to be
increased sharply~\cite{vara}. Basically, FE are highly
insulating, which prohibits direct current transport measurements.
Consequently, measurements on polarization and dielectric constant
are of more relevant to characterize FE. Due to its dielectric
effect, FE has been exploited in microwave devices such as tunable
oscillators, phase shifters and
varactors~\cite{mueller2,gevorgian3,korn4,apalkov}. Moreover, FE
also have the potential applications for detectors and devices
based on the energy conversion principle namely,
mechanical$\Leftrightarrow$thermal$\Leftrightarrow$electrical
energies~\cite{schubring5}.

Yet another technological advancement of FE is its functionality
as a gate insulator in Field-Effect Transistor (FET) as reported
extensively by de Boer {\it et al}.~\cite{boer6} and Podzorov {\it
et al}.~\cite{podzorov7} using organic molecules namely,
Tetracene, Pentacene and Rubrene. One of their interesting
observation is the dielectric constant's inverse dependence on
carrier mobility. Furthermore, developments of the dielectric gate
that consists of La, Al and Oxygen to substitute the common
SiO$_2$ has been successfully produced and even
patented~\cite{kaushik}. Perovskite FE with Ferromagnetic
properties such as (Tb,Y,Bi)MnO$_3$ have been investigated
experimentally~\cite{quezel8,kimura,huang9,seshadri10} due to its
possible technological impact on magneto-electric media, including
the Dynamic Random Access Memory (DRAM) devices. BST related
oxides are studied in the thin film form, usually grown by
pulsed-laser deposition technique. Parallel to this, there are
reports on the influence of the film's thickness with FE. In this
respect, a generalized Ginzburg-Devonshire theory was employed to
evaluate the behavior of the FE thin film coated with two metallic
electrodes~\cite{lu15}. Zheng~\textit{et al}.~\cite{lu15} found
that imperfect surface produces non-uniform polarization
distribution and subsequently gives rise to the phase transition
temperature or $T_C$. Apart from oxides, liquid crystal is one of
the well known soft matter that works on the principle of
molecular polarization~\cite{kaur11}. By applying an intense
alternating electric field during Triglycine Sulphate (TGS)
crystal growth, Arunmozhi~\textit{et al}.~\cite{arunmozhi12}
concluded that this procedure leads to a longer relaxation time,
which in turn reduces the dielectric constant and the spontaneous
polarization. On the theoretical aspect, a thermodynamical model
has been proposed to explain the $T$ and \textbf{E} dependence of
polarization and susceptibility of TGS by Otolinska~\textit{et
al}.~\cite{otolinska13}. Moreover, non-oxide alloy, TlInS$_2$ was
reported theoretically~\cite{mikailov14} to explain the
peculiarities in the dielectric susceptibility, which is connected
with the assumption that there exists coexistence of proper and
improper FE, somewhat mimicking the TbMnO$_3$, which also
demonstrates the said coexistence~\cite{kimura}.

Simple non-Fermi gas system and strongly correlated matter are
known to exhibit remarkable electronic properties with minuscule
substitutional doping. Handling such systems theoretically with
the ionization energy ($\mathcal{E_I}$) based Fermi-Dirac
statistics (iFDS) has been shown to be precise. For example,
simple systems such as the diluted magnetic semiconductors,
Mn$_{0.02}$Ge$_{0.98}$~\cite{arulsamy28} and
Ga$_{1-x}$Mn$_x$As~\cite{arulsamy29,arulsamy34} as well as for
complex crystals namely, Cuprate high-$T_c$ superconductors,
YBa$_2$Cu$_3$O$_7$~\cite{arulsamy30,arulsamy31,arulsamy32,arulsamy33,arulsamy,andrew}
and Manganite ferromagnets,
La$_{1-x}$Ca$_x$MnO$_3$~\cite{arulsamy29,arulsamy34} have been
shown quantitatively to be within the scope of iFDS. Therefore,
iFDS is further applied here to scrutinize theoretically the
evolution of polarization, static dielectric constant and the
screened Coulomb potential with doping in FE BST and its
derivatives. The reasons to revivify the dielectric functions and
the screened Coulomb potential with iFDS is two-fold. The first
one is as stated above, while the second reason being, to further
our evaluation on the actual role of phonon in the Cuprate
superconductors and also to justify why hole compensation effect
is unnecessarily invoked in the First-Principles approach for the
ferromagnetic semiconductors, both of which will not be addressed
here. It is well known that for free-electrons, the Thomas-Fermi
screening length determines the screened Coulomb potential while
the bare Coulomb potential is recovered when the electrons density
is literally zero. However, it is highlighted here that the bare
Coulomb potential can still be recovered in the presence of finite
electrons density, provided that $\mathcal{E}_I$ approaches
$\infty$. In addition to the theoretical justifications,
experimental results from Pauling~\cite{pauling35}, Jaswal {\it et
al}.~\cite{jaswal36} and Bell {\it et al}.~\cite{bell37} will be
used to further reinforce the accuracy of iFDS. Unfortunately, the
theory presented here is not yet suitable to predict the $T$
dependence of dielectric constant and polarizability, rather the
substitutional doping dependence of the mentioned parameters will
be accentuated and analyzed in detail.

\section{2. Theoretical details}\lb{s-in}

\subsection{2.1. Polarizability and optical dielectric function}\lb{s-in}

The electronic polarization, \textbf{P} of a particular crystal
can be defined as

\begin {eqnarray}
&\textbf{P}& = \mathcal{C}\sum_j \alpha_j(n_j)
\textbf{E}^{local}_j. \label{eq:1}
\end {eqnarray}

$\mathcal{C}$, $\alpha_j(n_j)$ and \textbf{E}$^{local}_j$ denote
the atoms concentration, polarizability at atom $j$, which is
proportional to electrons number in atom $j$, and the local
electric field at atom site $j$ respectively~\cite{kittel38}. The
local electric field for individual ions can be written as
\textbf{E}$_{local} = \textbf{E} +
\frac{1}{3\epsilon_0}\textbf{P}$ by assuming a spherical geometry
for individual ions and the crystals are purely
ionic~\cite{tessman39}. \textbf{E} is the macroscopic electric
field while $\frac{1}{3\epsilon_0}$ is the Lorentz factor. In
order to understand the evolution of \textbf{P} with $n$ as a
result of doping, another equation that relates doping with $n$ is
required. To this end, iFDS is utilized that eventually gives

\begin{eqnarray}
&n& = \int\limits^{\infty}_0{f_e(E)N_e(E)dE}, \nn \\&& =
\frac{1}{2\pi^2}\bigg(\frac
{2m_e}{\hbar^2}\bigg)^{3/2}e^{\lambda(E_F^0 - \mathcal{E}_I)}
\int\limits_0^\infty E^{1/2} \exp(-\lambda E) dE \nn
\\&& = 2\bigg(\frac{m_e}{2\lambda\pi\hbar^2}\bigg)^{3/2}
\exp\big[\lambda(E_F^0 - \mathcal{E_I})\big]. \label{eq:2}
\end{eqnarray}

Here, $\mu = -\lambda E_F^0$ whereas $\mathcal{E_I}$ represents
the electron-ion Coulomb attraction, which is also an accurate
representation of
doping-parameter~\cite{arulsamy29,arulsamy30,arulsamy31,arulsamy33,arulsamy34}.
The derivation of iFDS, $f(\mathcal{E_I}) =
\exp\big[-\mu-\lambda(E_{initial~state} \pm \mathcal{E_I})\big]$
by employing the restrictive conditions, $\sum_i^{\infty} n_i = n$
and $\sum_i^{\infty} (E_{initial~state}\pm \mathcal{E_I})_i n_i =
E$ can be found in the
Refs.~\cite{arulsamy30,arulsamy31,arulsamy32,arulsamy33,arulsamy34}.
$E_{initial~state}$ denotes the energy at certain initial state
and the Lagrange multipliers, $\mu_e + \lambda \mathcal{E_I} =
-\ln\big[(n/V)(2\pi\lambda\hbar^2/m_e)^{3/2}\big]$ and $\mu_h -
\lambda \mathcal{E_I} =
\ln\big[(p/V)(2\pi\lambda\hbar^2/m_h)^{3/2}\big]$. $V$ is the
volume in \textbf{k}-space and $E_F^0$ is the Fermi level at $T$ =
0 K. $m_{e,h}$ is the electrons or holes mass, $\hbar = h/2\pi$,
$h$ is the Planck constant, while $n$ and $p$ are the electrons
and holes number. In the previous
work~\cite{arulsamy33,arulsamy34}, $\lambda$ = $1/k_BT$ since the
variation of $n$ was with respect to the variation of $T$.
Contrary to the previous objective, $\lambda$ in this case can be
connected with the binding energy of an electron in a Hydrogen
atom, which is given by $E_{\verb"n"} =
(m/2\hbar^2)(e^2/4\pi\epsilon_0)^2(1/\verb"n"^2)$. Here,
$\verb"n"$ denotes the principal quantum number. The reason is to
calculate the variation of $n$ due to electrostatic potential, of
which the $n$ as given in Eq.~(\ref{eq:2}) is actually the induced
electrons density, $n_{ind}$ arises as a result of both external
and induced electrostatic potential. To obtain such a result, one
has to make use of the two restrictive conditions introduced in
iFDS as stated above that respectively give

\begin{eqnarray}
&&n = \frac {V}{2\pi^2}e^{-\mu - \lambda \mathcal{E}_I}
\int\limits_0^\infty {\bf k}^2 \exp\left(-\lambda
\frac{\hbar^2{\bf k}^2}{2m}\right) d{\bf k}, \nn
\\&& E = \frac {V\hbar^2}{4m\pi^2}  e^{-\mu -\lambda \mathcal{E}_I}
\int\limits_0^\infty {\bf k}^4 \exp\left(-\lambda\frac{\hbar^2{\bf
k}^2}{2m}\right)d{\bf k}.\nn
\end{eqnarray}

Solving the above integrals, one can arrive at $E = 3n/2\lambda =
E_{\verb"n"}$, hence $\lambda =
(12\pi\epsilon_0/e^2)\verb"n"^2r_B$, after taking $n$ = 1. $r_B$
represents the Bohr radius. Notice the essential factor,
$e^{\lambda(E_F^0 - \mathcal{E}_I)}$ in Eq.~(\ref{eq:2}) that will
be used together with $\lambda$ to define the carrier density
above $E_F^0$ as a result of the external and induced
electrostatic field and finally to explicitly derive the static
dielectric function and the screened Coulomb potential for the
strongly correlated (non-Fermi gas) systems. The dielectric
function ($\epsilon(\omega,\textbf{k})$) for an isotropic
polarization can be defined as~\cite{kittel38}

\begin {eqnarray}
&&\lim_{\textbf{k}\rightarrow 0} \epsilon(\omega) =
\frac{1}{\textbf{E}}\bigg[\textbf{E} +
\frac{\mathcal{C}}{\epsilon_0}\sum_j \alpha_j(n_j)
\textbf{E}^{local}_j\bigg]. \nn \\&& \epsilon(\omega,0) = 1 +
\frac{\mathcal{C}}{\textbf{E}\epsilon_0}\sum_j \alpha_j(n_j)
\textbf{E}^{local}_j. \label{eq:3}
\end {eqnarray}

$\epsilon(\omega = 0)$ and $\epsilon(\omega = \infty)$ are the
static and optical dielectric constants respectively, while
$\omega$ is the frequency. Again, one can easily notice the
proportionality, $\epsilon(0)$ $\propto$ $\alpha_j(n_j)$ $\propto$
$e^{\lambda(E_F^0 - \mathcal{E}_I)}$ without involving much
triviality. In fact, the parameters, $\alpha_j(n_j)$ and
$\epsilon(0)$ will be analyzed with respect to $\mathcal{E_I}$ at
constant $T$ and its correctness will be justified with the static
dielectric function and also with experimental results.

\subsection{2.2. Static dielectric function}\lb{s-in}

If the positive ions in the background are allowed to have a
sinusoidal variation that leads to the positive charge density,
$\rho^+(x)$ = $n^+_0e$ + $\rho_{ext}(\textbf{k})\sin
(\textbf{k}x)$ in the $x$ direction, then the response of the
electrons charge density is $\rho^-(x)$ = $-n^-_0e$ +
$\rho_{ind}(\textbf{k})\sin (\textbf{k}x)$. The term,
$\rho_{ext}(\textbf{k})\sin (\textbf{k}x)$ defines the external
electrostatic field that acts on the electrons. In other words,
the electrons will be deformed as a result of
$\varphi_{ext}(\textbf{k})$ and $\varphi_{ind}(\textbf{k})$. The
latter being the induced electrostatic potential. One can employ
the Poisson equation, $\triangledown^2\varphi =
-\frac{\rho}{\epsilon_0}$, $\varphi(\textbf{k})$ =
$\varphi_{ext}(\textbf{k})$ + $\varphi_{ind}(\textbf{k})$ and
$\rho(\textbf{k})$ = $\rho_{ext}(\textbf{k})$ +
$\rho_{ind}(\textbf{k})$ in order to arrive at~\cite{kittel38}

\begin {eqnarray}
&\varphi(\textbf{k})& =
\frac{\rho(\textbf{k})}{\textbf{k}^2\epsilon_0}. \label{eq:4}
\end {eqnarray}

The FE here concerns with the displacement of charge
(\textbf{D}(\textbf{k})), thus $\epsilon(0,\textbf{k})$ can be
defined~\cite{kittel38} as \textbf{D}(\textbf{k}) =
$\epsilon$(0,\textbf{k})\textbf{E}(\textbf{k}). Using,
$\triangledown \cdot \textbf{D} = \triangledown \cdot \epsilon
\textbf{E}$ = $\rho_{ext}/\epsilon_0$, $\triangledown \cdot
\textbf{E} = \rho/\epsilon_0$ and the Poisson equation, one can
also show that~\cite{kittel38}

\begin {eqnarray}
&\epsilon(0,\textbf{k})& =
\frac{\int^{\infty}_{-\infty}\rho_{ext}(\textbf{k})e^{i\textbf{k}
\cdot
r}d\textbf{k}}{\int^{\infty}_{-\infty}\rho(\textbf{k})e^{i\textbf{k}
\cdot r}d\textbf{k}} =
\frac{\int^{\infty}_{-\infty}\varphi_{ext}(\textbf{k})e^{i\textbf{k}
\cdot
r}d\textbf{k}}{\int^{\infty}_{-\infty}\varphi(\textbf{k})e^{i\textbf{k}
\cdot r}d\textbf{k}}. \nn \\&& \label{eq:5}
\end {eqnarray}

Now, at absolute zero (0 K), the Fermi level is given by $E_F^0 =
\frac{\hbar^2}{2m_e}(3\pi^2n_0)^{2/3}$. However, in the presence
of $\varphi_{ext}(x)$ and $\varphi_{ind}(x)$, $E_F(x) =
\frac{\hbar^2}{2m_e}[3\pi^2n(x)]^{2/3}$. Using the the
Thomas-Fermi (TF) approximation, one can write $E_F(x) - E_F^0 =
e\big[\varphi_{ext}(x) + \varphi_{ind}(x)\big] = e\varphi(x)$. As
a consequence, $\mathbb{L}$inearization can be carried out to
estimate the induced carrier density, $n(x) - n_0$ at certain
point $x$, in which $E_F^0(n_0)$ can be linearized as
$\mathbb{L}(n_0$) at $n_0 = n(x)$ to give

\begin {eqnarray}
\mathbb{L}(n_0) = E_F(x) + \frac{dE_F^0}{dn_0}\big[n_0 -
n(x)\big].\label{eq:6}
\end {eqnarray}

Since the standard linear approximation gives $\mathbb{L}(n_0)
\approx E_F^0$ and $dE_F^0/dn_0 = 2E_F^0/3n_0$, one can rewrite
Eq.~(\ref{eq:6}) as given below

\begin {eqnarray}
&&\frac{dE_F^0}{dn_0}\big[n(x) - n_0\big] \cong E_F(x) - E_F^0
 \cong
e\varphi(x).\nonumber
\\&& n(x) - n_0 \cong \frac{3n_0}{2E_F^0}e\varphi(x). \label{eq:7}
\end {eqnarray}

Utilizing iFDS or Eq.~(\ref{eq:2}) specifically, such that $n =
n_{ind} = n(x) - n_0$, then Eq.~(\ref{eq:7}) can be expressed as

\begin {eqnarray}
n = n(x) - n_0 \cong
\frac{3n_0}{2E_F^0}e\varphi(x)\exp\big[\lambda(E_F^0-\mathcal{E_I})\big].
\label{eq:8}
\end {eqnarray}

Consequently, using $\lim_{\mathcal{E}_I \rightarrow \infty} n(x)
= n_0$, one can surmise that it is rather impossible to induce a
region of enhanced electron concentration. However,
$\lim_{\mathcal{E}_I \rightarrow E_F^0} n(x) = n_0\big[1 +
(3/2E_F^0)e\varphi(x)\big]$ supports the free-electron or
Boltzmann-particle systems, which also satisfies the original TF
approximation. It is worth noting that, for the latter limit, the
electrons with its total energy, $TE$ $>$ $E_F^0$ are literally
free, forming the Fermi gas. In contrast, Eq.~(\ref{eq:8}) denies
such scenario, as long as $E_F^0 < TE < \mathcal{E}_I$. To see
this effect clearly, the screened Coulomb potential is derived and
discussed in the subsequent section. Employing Eq.~(\ref{eq:4})
and $\rho_{ind}(x) = -\big[n^-(x) - n_0^-\big]e$ (as introduced
earlier), the Fourier components of Eq.~(\ref{eq:8}) can be
written as

\begin {eqnarray}
&&\int\limits^{\infty}_{-\infty}\rho_{ind}(\textbf{k})e^{i\textbf{k}
\cdot r}d\textbf{k} \nn
\\&&
=
-\frac{3n_0}{2E_F^0}e^2\exp\big[\lambda(E_F^0-\mathcal{E_I})\big]
\int\limits^{\infty}_{-\infty}\varphi(\textbf{k})e^{i\textbf{k}
\cdot r}d\textbf{k} \nonumber \\&& = -\frac{3n_0e^2}{\epsilon_0
2E_F^0}\exp\big[\lambda(E_F^0-\mathcal{E_I})\big]
\int\limits^{\infty}_{-\infty}\frac{\rho(\textbf{k})}{\textbf{k}^2}e^{i\textbf{k}
\cdot r}d\textbf{k}. \label{eq:9}
\end {eqnarray}

Using Eq.~(\ref{eq:5}) with term by term division, the Fourier
component of the static dielectric function at zero frequency,
$\epsilon(0,\textbf{k})$ can be derived from Eq.~(\ref{eq:9}) as
given below

\begin {eqnarray}
\epsilon(0,\textbf{k}) = 1 +
\frac{\mathcal{K}_s^2}{\textbf{k}^2}\exp\big[\lambda(E_F^0-\mathcal{E_I})\big].\label{eq:10}
\end {eqnarray}

Firstly, notice that both $\epsilon(0,\textbf{k})$ and
$\epsilon(\omega,0)$ are proportional to the factor of
$\exp\big[\lambda(E_F^0-\mathcal{E_I})\big]$ and
$\epsilon(0,\textbf{k}) \neq \epsilon(\omega,0)$. Secondly, the TF
screening parameter remains the same, $\mathcal{K}_s^2 =
3n_0/2\epsilon_0E_F^0e^2$. Considering the Fermi-gas system with
$E_F^0 = \mathcal{E}_I$, then one can arrive at the original TF
dielectric function, $\epsilon(0,\textbf{k}) = 1 +
\mathcal{K}_s^2/\textbf{k}^2$. If however, $\mathcal{E}_I
\rightarrow \infty$, then $\epsilon(0,\textbf{k}) \rightarrow
constant$.

\subsection{2.3. Screened Coulomb potential}\lb{s-in}

Imagine a test charge $e$ is placed at certain energy level
($>E_F^0$) in the background of a sinusoidal variation of the
positive ions, then the Fourier transformed electrostatic
potential for the unscreened Coulomb potential,
$\varphi_{q}(\textbf{k})$ can be shown to be
$q/\textbf{k}^2\epsilon_0$ using Eq.~(\ref{eq:4}). The Fourier
inversion of $\varphi_{q}(\textbf{k})$ is $\varphi_{q}(r) =
\big[1/(2\pi)^3\big] \int_0^\infty
2\pi\textbf{k}^2(q/\epsilon_0\textbf{k}^2)e^{i\textbf{k}\cdot r}
d\textbf{k} = \big[q/4\pi^2\epsilon_0\big] \int_0^\infty
d\textbf{k} \int_{-1}^1 e^{i\textbf{k}r\cos \theta} d(\cos \theta)
= q/4\pi\epsilon_0r$ as it should be. Subsequently, one can derive
the screened Coulomb potential, $\varphi(r)$ using
Eq.~(\ref{eq:5}), in which $\epsilon(0,\textbf{k}) =
\varphi_{q}(\textbf{k})/\varphi(\textbf{k})$. As such,
$\varphi(\textbf{k}) =
(q/\epsilon_0\textbf{k}^2)\big[\textbf{k}^2/(\textbf{k}^2 +
\mathcal{K}_{s,I}^2)\big]$. Notice the parameterization,
$\mathcal{K}_{s,I}^2 =
\mathcal{K}_s^2\exp\big[\lambda(E_F^0-\mathcal{E_I})\big]$ that
has been used for simplicity. Finally, the Fourier inversion of
$\varphi(\textbf{k})$ is $\varphi(r)$ that can be derived as

\begin {eqnarray}
\varphi(r) &&= \frac{1}{(2\pi)^3} \int\limits_{-\infty}^\infty
\frac{\pi\textbf{k}^2q}{\epsilon_0(\textbf{k}^2 +
K_{s,I}^2)}d\mathbf{k} \nn\\&& \times \int\limits_{-1}^1
e^{i\textbf{k}r\cos \theta} d(\cos \theta) \nn \\&& =
\frac{q}{2\pi^2\epsilon_0r}\int\limits_0^\infty
\frac{\textbf{k}d\textbf{k}}{\textbf{k}^2 + \mathcal{K}_{s,I}^2}
\sin(\textbf{k}r) \nn\\&& =
\frac{q}{4\pi\epsilon_0r}\exp\big[-\mathcal{K}_sre^{\frac{1}{2}\lambda(E_F^0-\mathcal{E_I})}\big].\label{eq:11}
\end {eqnarray}

$\pi\textbf{k}^2$ denotes the area in \textbf{k}-space. The
screened Coulomb potential in this case is a function of doping
parameter, $\mathcal{E}_I$ and one can enumerate the evolution of
$\varphi(r)$ with doping for non-Fermi gas systems. The
$\mathcal{E}_I$ that originates from iFDS can be used to justify
that an electron to occupy a higher state $N$ from the initial
state $M$ is more probable than from the initial state $L$ if the
condition $\mathcal{E}_I(M)$ $<$ $\mathcal{E}_I(L)$ at certain $T$
is satisfied. As for a hole to occupy a lower state $M$ from the
initial state $N$ is more probable than to occupy the state $L$ if
the same condition above is satisfied.
$\mathcal{E}_{initial~state}$ is the energy of a particle in a
given system at a certain initial state and ranges from $+\infty$
to 0 for the electrons and 0 to $-\infty$ for the holes. Simply
put, the magnitude of $\varphi(r)_{M \leftrightarrow N}$ due to
the polarization initiated excitation of an electron (\verb"hole")
from the initial state $M$ ($N$) to $N$ ($M$) is quantitatively
smaller than $\varphi(r)_{L \leftrightarrow N}$. This effect can
be non-trivially verified from the Eq.~(\ref{eq:11}). On the
contrary however, placing a test charge with its $TE$ $>$ $E_F^0$
and $E_F^0 = \mathcal{E}_I$ imply Fermi gas that eventually gives
rise to the TF screening potential~\cite{kittel38} given by
$\frac{q}{4\pi\epsilon_0r}e^{-\mathcal{K}_sr}$. Add to that,
Eq.~(\ref{eq:11}) also reduces to the unscreened Coulomb potential
even at finite $n_0$ provided that $\mathcal{E}_I \rightarrow
\infty$. The limit, $\mathcal{E}_I \rightarrow \infty$
conceptually means that the electrons in the presence of the
sinusoidal potential of positive ions are infinitely rigid with
zilch polarization and eventually can be thought of as a sphere of
a net charge $q$, as seen by the previously placed test charge
$e$. Therefore, the potential is indeed a bare Coulomb potential,
$q/4\pi\epsilon_0r$. Refer to Fig.~\ref{fig1} to clearly observe
the transition of the TF screened potential ($\bullet$) to the
Coulomb bare potential ($\blacksquare$) as a result of iFDS, at
finite carrier density ($n_0 \neq 0$). The solid lines, $\bullet$
($\mathcal{E}_I \rightarrow E_F^0$) and $\blacksquare$
($\mathcal{E}_I \rightarrow \infty$) are obtained from
Eq.~(\ref{eq:11}) with its appropriate limits.

\section{3. Discussion}\lb{s-in}

Having derived all the required functions, now it is possible to
venture into the experimental data reported by various researchers
starting from the 1920s. Notice that the above derivations
strictly requires substitutional doping that significantly
maintains a single phase or solid solution. Obviously, doping with
a purpose of creating a second phase or multi-phases are not
applicable with iFDS based theory. The Clausius-Mossotti relation
in the optical range is given by~\cite{kittel38}

\begin{eqnarray}
\frac{\epsilon - 1}{\epsilon + 2} = \frac{\mathcal{N}^2 -
1}{\mathcal{N}^2 + 2} = \frac{\mathcal{C}}{3\epsilon_0}\sum_j
\alpha_j(n_j). \label{eq:12}
\end{eqnarray}

$\mathcal{N}$ represents the refractive index. Using this
relation, the electronic polarizability of atom $j$, which is a
function of electron number in atom $j$ ($\alpha_j(n_j)$) for a
wide variety of ions~\cite{pauling35,jaswal36} were calculated and
tabled in the Ref.~\cite{kittel38}. The magnitude of
$\sum_j\alpha_j(n_j)$ can be assumed to be accurate since those
ions (plasma) easily satisfy the Lorentz
factor~\cite{pauling35,tessman39} for spherical ions or atoms.
Figure~\ref{fig2} a)-d) depict the relation between
$\alpha_j(n_j)$ and $\mathcal{E}_I$ for 1+ $\rightarrow$ 4+ ions
respectively. Those experimental data
points~\cite{pauling35,jaswal36} were fitted in
Fig.~\ref{fig2}a)-d) with $\sum_j\alpha_j(n_j) =
\mathcal{X}\exp[\mathcal{Y}]$, which is in accordance with the
principle of Eqs.~(\ref{eq:1}),~(\ref{eq:2}) and~(\ref{eq:3}).
Here, the fitting parameters, $\mathcal{X}$ = constant of
proportionality while $\mathcal{Y}$ =
$\lambda(E_F^0-\mathcal{E}_I)$. Both values ($\mathcal{X}$ and
$\mathcal{Y}$), obtained from the fittings are given in the
Fig.~\ref{fig2}a)-d) itself. In accordance with iFDS, polarizing a
4+ ion, say Ce$^{4+}$ needs to overcome an energy proportional to
6325 kJmol$^{-1}$ (the 5$^{th}$ ionization energy), while
La$^{3+}$, Ba$^{2+}$ and Cs$^{+}$ need to overcome the energies
proportional to 4819 kJmol$^{-1}$ (the 4$^{th}$ ionization
energy), 3600 kJmol$^{-1}$ (the 3$^{rd}$ ionization energy) and
2234.3 kJmol$^{-1}$ (the 2$^{nd}$ ionization energy) respectively.
The absolute values are actually equal to the energy needed to
ionize an atom or ion such that the electron is excited to a
distance $r$. However, the ionization energies stated above are
for taking that particular electron to $ r \rightarrow \infty$.
Considering this scenario, one can surmise that both $\mathcal{X}$
and $\mathcal{Y}$ are predicted to reduce with increasing valence
state. As anticipated, the decreasing magnitudes of both
$\mathcal{X}$ and $\mathcal{Y}$ with increasing valence states
have been calculated so as to fit the experimental data points.
Furthermore, one can also understand the exponential decrease of
$\alpha_j(n_j)$ with 1+ (Fig.~\ref{fig2}a)), 2+
(Fig.~\ref{fig2}b)), 3+ (Fig.~\ref{fig2}c)) and 4+
(Fig.~\ref{fig2}d)) ions, which are as a result of increased
$\mathcal{E}_I$. Simply put, the exponential reduction of
$\alpha_j(n_j)$ for Cs$^{+}$ $\to$ ... Li$^{+}$, Ba$^{2+}$ $\to$
... Be$^{2+}$, La$^{3+}$ $\to$ ... B$^{3+}$ and Ce$^{4+}$ $\to$
... C$^{4+}$ are due to increased $\mathcal{E}_I$. Similarly, the
reduction of both $\mathcal{X}$ and $\mathcal{Y}$ is also due to
reduced $\alpha_j(n_j)$ for high valence state ions (4+) as
compared with 1+ ions. It is important to realize that
$\mathcal{Y}$ $\propto$ $\lambda$, which implies that
$\mathcal{Y}$ increases with increasing ionic size since $\lambda
\propto \verb"n"^2r_B$ as defined earlier.

Out of naivety, one should not assume that $\mathcal{Y}$ is
supposed to increase from Li$^{+}$ $\rightarrow$ Be$^{2+}$
$\rightarrow$ B$^{3+}$ $\rightarrow$ C$^{4+}$ since the ionic size
is increasing. This is because, different elements have different
ionic properties due to different number of positive charge
protons, which determines the Coulomb interactions. Actually,
$\mathcal{Y}$ will increase from \{C, Si, Ti, Zr, Ce\}$^{4+}$
$\rightarrow$ \{C, Si, Ti, Zr, Ce\}$^{3+}$ $\rightarrow$ and so on
for the respective ions. The average $\mathcal{E}_I$ has been
determined with, $\mathcal{E}_I [X^{z+}] =
\sum_{i=1}^z\frac{\mathcal{E}_{Ii}}{z}$ where $z$ is the valence
state. Prior to averaging, the 1$^{st}$, 2$^{nd}$, 3$^{rd}$ and
4$^{th}$ ionization energies for all the elements mentioned above
were taken from Ref.~\cite{winter}. In summary, the
proportionality, $\alpha_j(n_j)$ $\propto$ $e^{\lambda(E_F^0 -
\mathcal{E}_I)}$ is indeed valid. On the other hand, the static
dielectric constant, $\epsilon(0,\textbf{k})$ must also satisfy
Eq.~(\ref{eq:10}). To this end, BaTiO$_3$ ($\bullet$), SrTiO$_3$
($\blacksquare$), CaTiO$_3$ ($\blacktriangle$) and
Ba$_{0.5}$Sr$_{0.5}$TiO$_3$ ($\blacklozenge$) samples obtained
from Ref.~\cite{bell37} are plotted in Fig.~\ref{fig3} as a
function of average $\mathcal{E}_I$ at $T1 = T_C + 200$ K. The
average $\mathcal{E}_I$s for Ba$^{2+}$, Sr$^{2+}$ and Ca$^{2+}$
are given by 734, 807, and 867 kJmol$^{-1}$ respectively. After
applying iFDS, one can predict that $\epsilon(0,\textbf{k})$
should decrease from BaTiO$_3$ $\rightarrow$ SrTiO$_3$
$\rightarrow$ CaTiO$_3$ while the magnitude of
$\epsilon(0,\textbf{k})$ for Ba$_{0.5}$Sr$_{0.5}$TiO$_3$ is
expected to be between BaTiO$_3$ and SrTiO$_3$. All these
predictions are remarkably in accordance with Eq.~(\ref{eq:10}).
In fact, those experimental data~\cite{bell37} have been
reproduced with significant accuracy using Eq.~(\ref{eq:10}) as
indicated with a solid line. Equation~(\ref{eq:10}) has been
rewritten as $\epsilon(0,\textbf{k}) =
\mathcal{X}\exp[\mathcal{Y}]$ in which the fitted values for
$\mathcal{X}$ and $\mathcal{Y}$ are given in Fig.~\ref{fig3}
itself. Equation~(\ref{eq:10}) has been rewritten in the stated
form because $\epsilon(0,\textbf{k})$ $\gg$ 1 or
$\frac{\mathcal{K}_s^2}{\textbf{k}^2}\exp\big[\lambda(E_F^0-\mathcal{E_I})\big]$
$\gg$ 1 for BST ferroelectrics. As a consequence, one can write
that $\mathcal{X}$ = $\mathcal{K}_s^2/\textbf{k}^2$ and
$\mathcal{Y}$ remains the same as $\lambda(E_F^0-\mathcal{E}_I)$.
In other words, one can understand why the static dielectric
constant becomes smaller exponentially from BaTiO$_3$
$\rightarrow$ Ba$_{0.5}$Sr$_{0.5}$TiO$_3$ $\rightarrow$ SrTiO$_3$
$\rightarrow$ CaTiO$_3$, which is due to increasing
$\mathcal{E}_I$ ($\mathcal{E}_I$(Ba$^{2+}$) $<$
$\mathcal{E}_I$(Sr$^{2+}$) $<$ $\mathcal{E}_I$(Ca$^{2+}$)).

\section{4. Conclusions}\lb{s-in}

In conclusion, the evolution of polarizability, polarization,
optical-static dielectric constants and the screened Coulomb
potential has been derived for non-Fermi gas system using iFDS.
These functions are able to explain the doping effect as well as
reproduces the experimental doping trend with high accuracy in the
well known BST and its related Perovskites namely, BaTiO$_3$,
Ba$_{0.5}$Sr$_{0.5}$TiO$_3$, SrTiO$_3$ and CaTiO$_3$. The
relationship between polarizability and iFDS is also found to be
highly precise as compared with Pauling's experimental data on 1+,
2+, 3+ and 4+ ions, which in turn justifies the applicability of
iFDS.

\section*{Acknowledgments}\lb{s-in}
The author is grateful to Arulsamy Innasimuthu, Sebastiammal
Innasimuthu, Arokia Das Anthony and Cecily Arokiam of CMG-A for
their hospitality, as well as for the financial support. ADA also
thanks Hendry Izaac Elim and Chong Kok Boon for providing some of
the references.

\begin{figure}
\caption {The screened electrostatic potential with iFDS effect
has been calculated against $r$ with Eq.~(\ref{eq:11}). The solid
lines are from Eq.~(\ref{eq:11}). In the limits, $\mathcal{E}_I
\rightarrow E_F^0$ and $\mathcal{E}_I \rightarrow \infty$, one can
recover the TF screened potential ($1/re^{-\mathcal{K}_sr}$) and
bare Coulomb potential ($1/r$) respectively. Notice that
$1/re^{-\mathcal{K}_sr}$ and $1/r$ are represented with filled
circle and filled square respectively.} \label{fig1}
\end{figure}

\begin{figure}
\caption {Polarizability as a function of electrons number,
$\alpha_j(n_j)$ are calculated (solid lines) against
$\mathcal{E}_I$ in order to fit the experimental data points for
a) 1+, b) 2+, c) 3+ and d) 4+ ions using $\sum_j\alpha_j(n_j) =
\mathcal{X}\exp[\mathcal{Y}]$, which is in accordance with the
principle of Eq.~(\ref{eq:3}). The fitting parameters,
$\mathcal{X}$ and $\mathcal{Y}$ are a constant of proportionality
and $\lambda(E_F^0-\mathcal{E}_I)$ respectively. All the
experimental data points given in a),b),c) and d) were obtained
from Pauling~\cite{pauling35} and Jaswal-Sharma~\cite{jaswal36}.}
\label{fig2}
\end{figure}

\begin{figure}
\caption {The experimental static dielectric constant,
$\epsilon(0,\textbf{k})$ at $T1 = T_C + 200$ K is fitted with
Eq.~(\ref{eq:10}) for BaTiO$_3$, SrTiO$_3$, CaTiO$_3$ and
Ba$_{0.5}$Sr$_{0.5}$TiO$_3$. Actually, Eq.~(\ref{eq:10}) has been
rewritten in the form of $\epsilon(0,\textbf{k}) =
\mathcal{X}\exp[\mathcal{Y}]$ (since $\epsilon(0,\textbf{k})$
$\gg$ 1 or
$\frac{\mathcal{K}_s^2}{\textbf{k}^2}\exp\big[\lambda(E_F^0-\mathcal{E_I})\big]$
$\gg$ 1), while its fitting parameters, $\mathcal{X}$ =
$\mathcal{K}_s^2/\textbf{k}^2$ and $\mathcal{Y}$ =
$\lambda(E_F^0-\mathcal{E}_I)$. The exponential decrease of
$\epsilon(0,\textbf{k})$ from BaTiO$_3$ $\rightarrow$
Ba$_{0.5}$Sr$_{0.5}$TiO$_3$ $\rightarrow$ SrTiO$_3$ $\rightarrow$
CaTiO$_3$ is associated with increasing $\mathcal{E}_I$ from
Ba$^{2+}$ $\rightarrow$ Sr$^{2+}$ $\rightarrow$ Ca$^{2+}$.}
\label{fig3}
\end{figure}

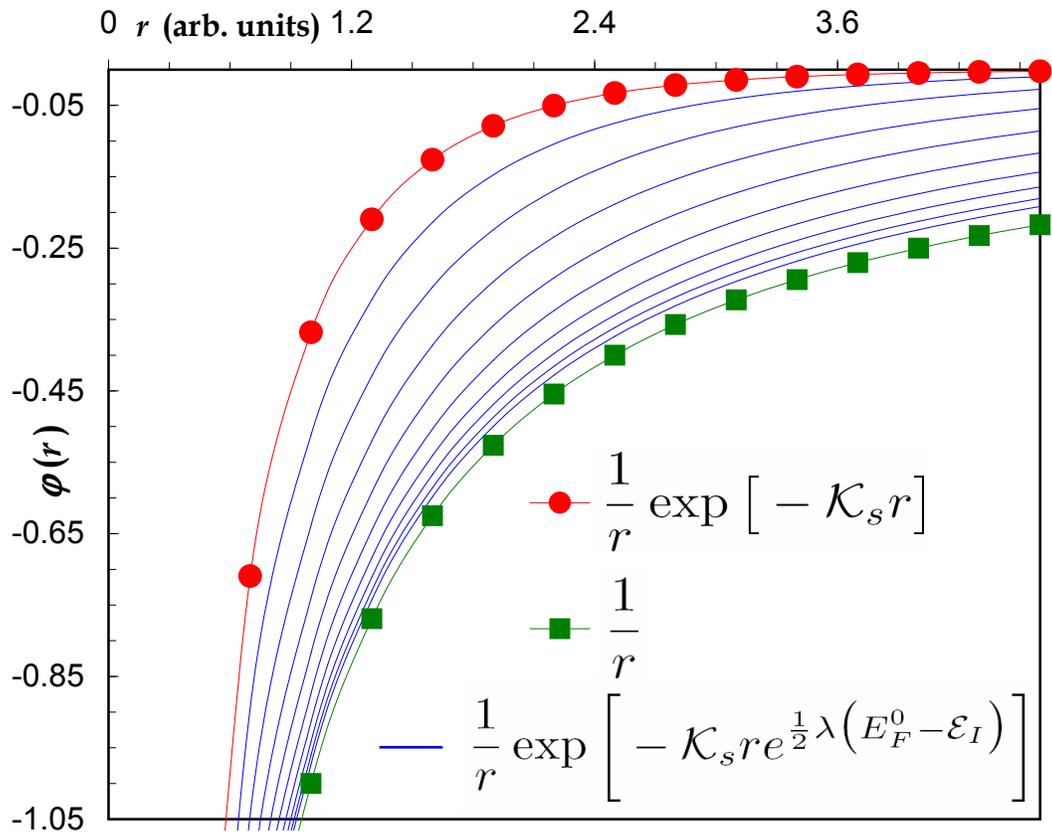
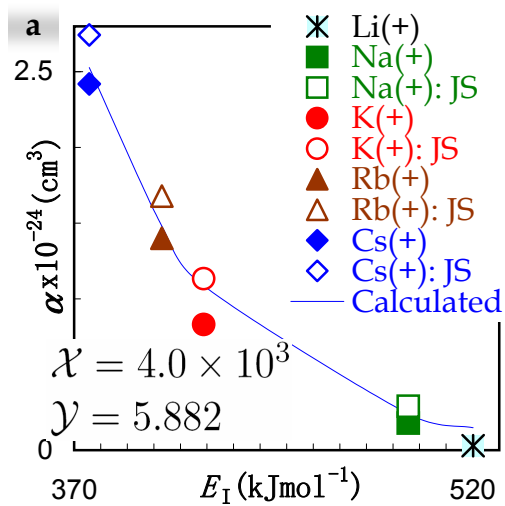
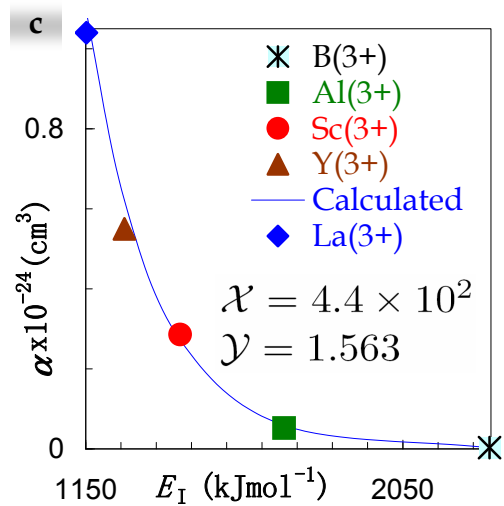
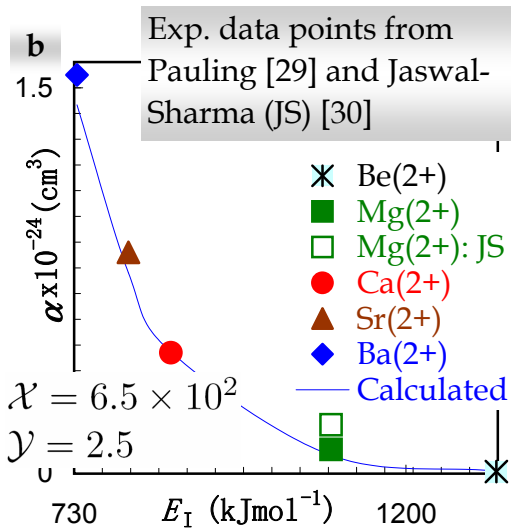
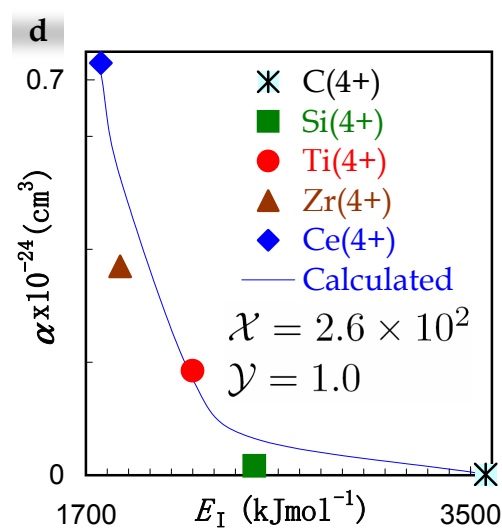
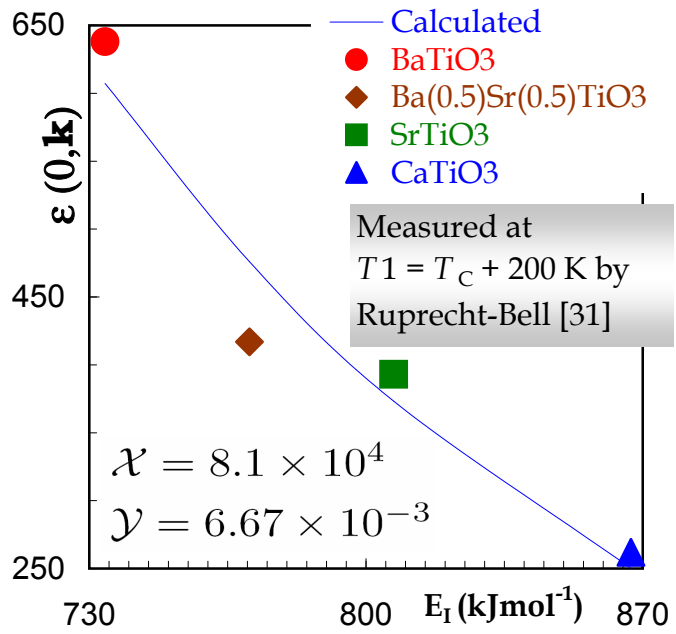
\end{document}